\documentclass{article}
\usepackage{lajolla2006}
\usepackage{graphicx}
\usepackage{floatflt,epsfig}
\usepackage{subfigure}
\usepackage{xspace}

\frompage{000} \topage{000}                                              

\newcommand{\lam}{\ensuremath{\Lambda} \xspace}

\newcommand{\kz}{\ensuremath{\mathrm{K^{0}_{S}}}\xspace}

\newcommand{\pt}{\ensuremath{p_{T}}\xspace}
\newcommand{\mt}{\ensuremath{m_{T}}\xspace}
\newcommand{\XI}{\ensuremath{\Xi^{-}}\xspace}

\newcommand{\pp}{\ensuremath{p+p}\xspace}
\newcommand{\ppbar}{\ensuremath{p+\bar{p}}\xspace}
\newcommand{\ee}{\ensuremath{e^{+}+e^{-}}\xspace}
\newcommand{\sqsRhic}{\ensuremath{\sqrt{s}=200}\xspace}
\newcommand{\sqsSps}{\ensuremath{\sqrt{s}=630}\xspace}

\def\rightmark{STAR strangeness data in \pp vs pQCD models}
\def\leftmark{M.Heinz (STAR)}

\title{``Can STAR \pp data help constrain fragmentation functions for strange hadrons''}
\authors{
{Mark T. Heinz (for the STAR collaboration)
}\\[2.812mm]
{\normalsize
Yale University, WNSL, 272 Whitney Avenue \\
CT 06517, New Haven, USA\\[0.2ex]
}}

\abstract{STAR has measured a variety of strange particle species in
\pp collisions at \sqsRhic GeV. These high statistics data are ideal
for comparing to existing leading- and next-to-leading order
perturbative QCD (pQCD) models. Leading-order (LO) models such as
PYTHIA need to be tuned to describe identified strange particle data
from STAR. We show that tuned PYTHIA can also describe the
\pt~spectra of strange resonances. More rigorous Next-to-Leading
order pQCD calculations using parameterized fragmentation functions
for quarks and gluons will also be compared to STAR data. The OPAL
experiment has recently released \ee data from light quark flavor
tagged analyses allowing for the first time to make precise
parameterizations of light flavor separated fragmentation function.
We show that our \lam~data put a more stringent constraint on the
gluon fragmentation function than \ee data. Furthermore we show that
pQCD fails to describe the observed enhancement of baryon-to-meson
ratio at intermediate \pt (2-6 GeV/c), which may be a first
indication of other, non-perturbative mechanisms at play in $p+p$
collisions at that momentum.}


\begin{document}

\maketitle

\setcounter{page}{1}

\section{Introduction}\label{intro}

Perturbative QCD has proven to be successful in describing inclusive
hadron production in elementary collisions. Within the theory's
range of applicability, calculations at next-to-leading order (NLO)
have produced accurate predictions for transverse momentum spectra
of inclusive hadrons at different energy scales
\cite{Borzumati:95,Marco:SQM04}. With the new high statistics
proton-proton data at \sqsRhic GeV collected by STAR, we can now
extend the study to identified strange hadrons as well as strange
resonances.

The perturbative QCD calculation applies the factorization ansatz to
calculate hadron production and relies on three ingredients. The
non-perturbative parton distribution functions (PDF) are obtained by
parameterizations of deep inelastic scattering data. They describe
quantitatively how the partons share momentum within a nucleus. The
second part, which is perturbatively calculable, consists of the
parton cross-section amplitude evaluated to LO or NLO using Feynman
diagrams. The third part consists of the non-perturbative
Fragmentation functions (FF) obtained from \ee collider data using
quark-tagging algorithms. These parameterized functions are
sufficiently well known for fragmenting light quarks, but less well
known for fragmenting gluons and heavy quarks. Recently, Kniehl,
Kramer and P\"otter (KKP) have shown that FF are universal between
\ee and \pp collisions \cite{KKP:01}.

In the following section we compare our p+p data to PYTHIA, the most
commonly used leading-order Monte Carlo event generator for
elementary collision. We then move on to compare our data with more
sophisticated NLO calculations. This will lead to a discussion of
the difference between quark and gluon jets in \pp. In the final
section we discuss the different contribution of these two jet types
to the production of baryons and mesons.

\section{Data Analysis}\label{analysis}

The present data were reconstructed using the STAR detector system
which is described in more detail elsewhere \cite{STAR2}. The main
detector used in this analysis is the Time Projection Chamber (TPC)
covering the full acceptance in azimuth and a large pseudo-rapidity
coverage ($\mid \eta \mid < 1.5$). A total of 14 million non-singly
diffractive (NSD) events were triggered with the STAR beam-beam
counters (BBC) requiring two coincident charged tracks at forward
rapidity. Due to the particulary low track multiplicity environment
in p+p collisions only 76\% of primary vertices are found correctly;
from the remainder, 14\% are lost and 10\% are badly reconstructed
as a MC-study showed. Of all triggered events, 7 million events
passed the selection criteria requiring a valid primary vertex
within 50cm along the beam-line from the center of the TPC detector.
The strange particles were identified from their weak decay to
charged daughter particles. The following decay channels and the
corresponding anti-particles were analyzed: $\mathrm{K^{0}_{S}}
\rightarrow \pi^{+} + \pi^{-}$ (b.r. 68.6\%), $\Lambda \rightarrow p
+ \pi^{-}$(b.r. 63.9\%) ,$\Xi^{-} \rightarrow \Lambda +
\pi^{-}$(b.r. 99.9\%). Particle identification of the daughters was
achieved by requiring the dE/dx to fall within the 3$\sigma$-bands
of the theoretical Bethe-Bloch parameterizations. Further background
in the invariant mass was removed by applying topological cuts to
the decay geometry. Corrections for acceptance and particle
reconstruction efficiency were obtained by a Monte-Carlo based
method of embedding simulated particle decays into real events and
comparing the number of simulated and reconstructed particles in
each $p_{T}$-bin.

\section{Comparison to PYTHIA}\label{pythia}

One of the most widely used models for simulating elementary
collisions is PYTHIA \cite{Pythia:87}. It is a parton-shower based
event generator that includes leading order parton processes and
parton fragmentation based on the Lund Model. The parton
distributions of the initial state protons can be chosen from an
array of PDFs (here we use CTEQ5M). The model is being actively used
and the authors have recently released a version with completely
overhauled multiple scattering and shower algorithms (version 6.3)
\cite{Pythia:04}. The PYTHIA version used in this paper is 6.317.

The string fragmentation based on the Lund Model requires only two
parameters to define the shape of the fragmentation function and is
universal for all light quark flavors. Baryons are produced from
di-quarks and their probability is suppressed with respect to
$\bar{q}q$ pair production. Next-to-leading order processes can be
"simulated" in PYTHIA by tuning the K-factor (MSTP(33)) or by
increasing the parton shower activity. This will enhance the
relative probability of hard processes of type quark-gluon and thus
mock-up the contributions from higher order processes. In figure 1
we first compare the measured STAR spectra for identified pions and
protons to a simulation from PYTHIA. The pions agree very well with
the default parameters whereas the protons seem to lie in between
the default and the tuned K-factor calculation.
\begin{figure}[h]
\begin{center}
\epsfig{figure=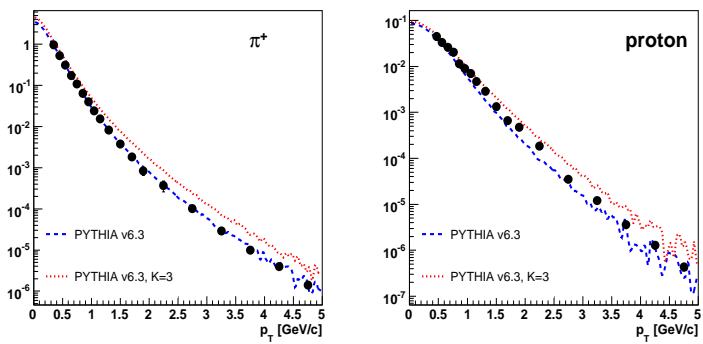, height=5cm,width=10cm}
\caption{Identified $\pi^{+}$ and protons in minimum-bias $p+p$
collisions at $\sqrt{s}$= 200 GeV compared to predictions from
PYTHIA v6.3 with and without K-factor. Data from \cite{Bedanga:06}}
\label{fig:nonstrange}
\end{center}
\end{figure}
In figure 2 (upper row) we compare PYTHIA calculations for strange
mesons and baryons to the measured STAR data. Whereas the default
parameters agree quite well for the \kz, they clearly underestimate
the yields at intermediate \pt for the \lam and \XI. By increasing
the K-factor to 3 we achieve a reasonable agreement with the data.
In figure \ref{fig:strangeLO} (lower row) we compare PYTHIA to the
strange resonances $K^{*}$, $\phi$ and $\Sigma^{*}$. Again, only
when applying a higher K-factor does the calculation agree with our
data.


\begin{figure}[ht]
\begin{center}
\epsfig{figure=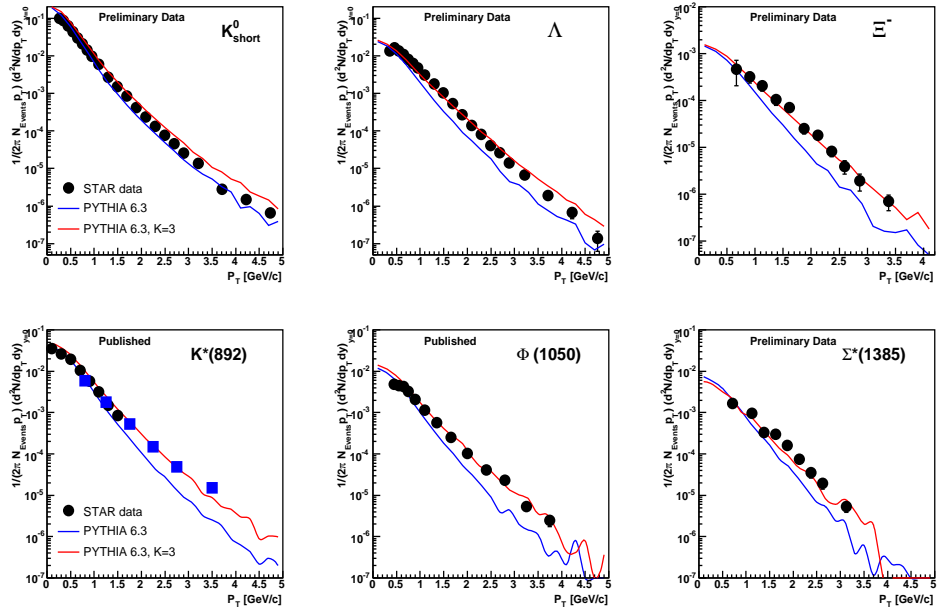, height=8.6cm,
width=13cm} \caption{(Top) Minimum-bias \pt~spectra for \kz, \lam
and $\Xi$ at ($\mid{y}\mid < 0.5$) from $p+p$ at $\sqrt{s} = 200
$GeV. (Bottom) $K^{*}$ and $\phi$, and $\Sigma^{*}$ \pt~spectra at
mid-rapidity. In the left panel blue squares are $K^{0*}$ and black
symbols are $K^{+*}$. Resonance data from
\cite{Kstar:05,Phi:05,Markert:06}. } \label{fig:strangeLO}
\end{center}
\end{figure}

In summary, PYTHIA is capable of describing \pt spectra for a
variety of particles from $p+p$ collisions at RHIC energies.
However, we have presented evidence that a tune of the LO K-factor
is necessary in particular for strange baryons and resonances. Of
course, we have not explored all possibilities of parameter ``tunes"
and there may be other, equivalent ways of reproducing the data.

What are the possible reasons for this discrepancy? The ``naive"
reason, supported by the K-factor tune, is that higher order
contribution may be significant. However it is troubling that the
pions in figure \ref{fig:nonstrange} do not seem to require this
tune, thus introducing a rather ``unnatural" particle species
dependance. Nevertheless, a similar study of K-factors for
non-identified hadrons found that at $\sqrt{s} = 200$GeV a value of
3 was needed \cite{Eskola:03}. Another, maybe more plausible
explanation may be related to fragmentation functions in PYTHIA. It
could be that some flavor dependant refinements to the Lund
symmetric string fragmentation are necessary. This will be discussed
in the next section with the use of NLO calculation and a set of
parameterized FF.

\section{Comparison to next-to-leading order pQCD}\label{nlo}

A next-to-leading order calculation can help solve both these
problems by including higher order parton processes \textbf{and}
rigorously parameterized fragmentation functions. Fragmentation
functions for separated quark flavors have been notoriously
difficult to obtain due to the lack of sufficiently precise collider
data. However, recently OPAL has published flavor tagged data which
allowed theorists to compute better fragmentation functions
\cite{OPAL:00}.

\begin{figure}[ht]
\begin{center}
\epsfig{figure=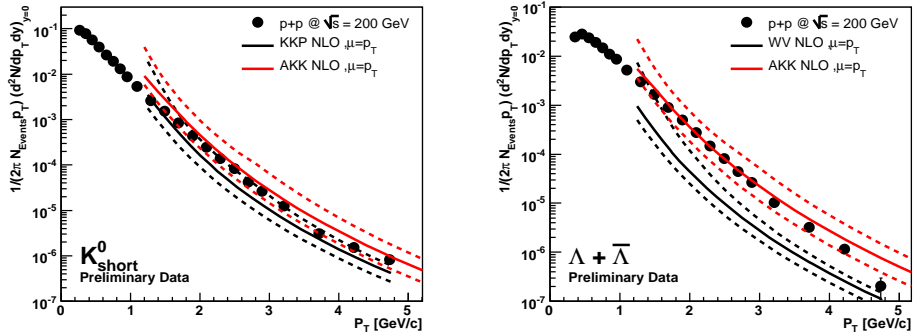, height=5cm, width=13cm}
\caption{\pt~spectra for \kz (left), \lam (right) at midrapity
($\mid{y}\mid < 0.5$) from $p+p$ at $\sqrt{s} = 200 $GeV compared to
two different NLO calculations. Dashed lines indicate the scale
uncertainty of the NLO calculation, ie. $\mu=0.5\pt$ (lower),
$\mu=2\pt$ (upper). } \label{fig:strangeNLO}
\end{center}
\end{figure}

In figure \ref{fig:strangeNLO} we compare two different NLO
calculations to our \kz and \lam data . The first one (black lines)
uses older FF by Kniehl \emph{et al.} (KKP) and Vogelsang \emph{et
al} (WV) \cite{deFlorian:PRD57}. The second one (red lines) was done
by Albino \emph{et al.} (AKK) using more recent FF based on light
flavor tagged OPAL data \cite{AKK:06}. Clearly, these newer
parameterizations improve the description of our \lam data greatly.
However, in order to achieve this agreement they fix the initial
gluon to \lam fragmentation function ($D_{g}^{\Lambda}$) to that of
the proton, then estimate that a additional scaling factor of 3 is
necessary to achieve agreement with STAR data. This modified FF for
$D_{g}^{\Lambda}$ however also works well in describing the
$p+\bar{p}$ SPS data at $\sqrt{s} = 630$GeV. So it appears that the
STAR data is a better constraint for the high z part of the gluon
fragmentation function than the OPAL \ee data. Similar conclusions
with respect to the important role of \pp collisions have been drawn
elsewhere \cite{Levai:PRL02}.

\section{Hadrons from quark vs gluon-jets}\label{quarkgluon}

In order to investigate the fragmentation of gluons further, we
looked at other observables that may be sensitive to this. For this
study we used the PYTHIA Monte Carlo generator to differentiate
events with gluons jets vs. quarks jets in the final state. We
define a ``Gluon-jet" event as one where the final partons are g-g
or g-q and a ``Quark-jet" event one where the final partons are q-q.
The overall weighting of these events in the total \pp cross-section
is dominated in favor of gluon-jets in PYTHIA. We then compared
PYTHIA to the arbitrarily scaled \mt~spectra from data. The scaling
was adjusted such that all spectra would overlap in the 0-1.5 GeV
region.

\begin{figure}[ht]
\centering \mbox{\subfigure[Scaled
data]{\epsfig{figure=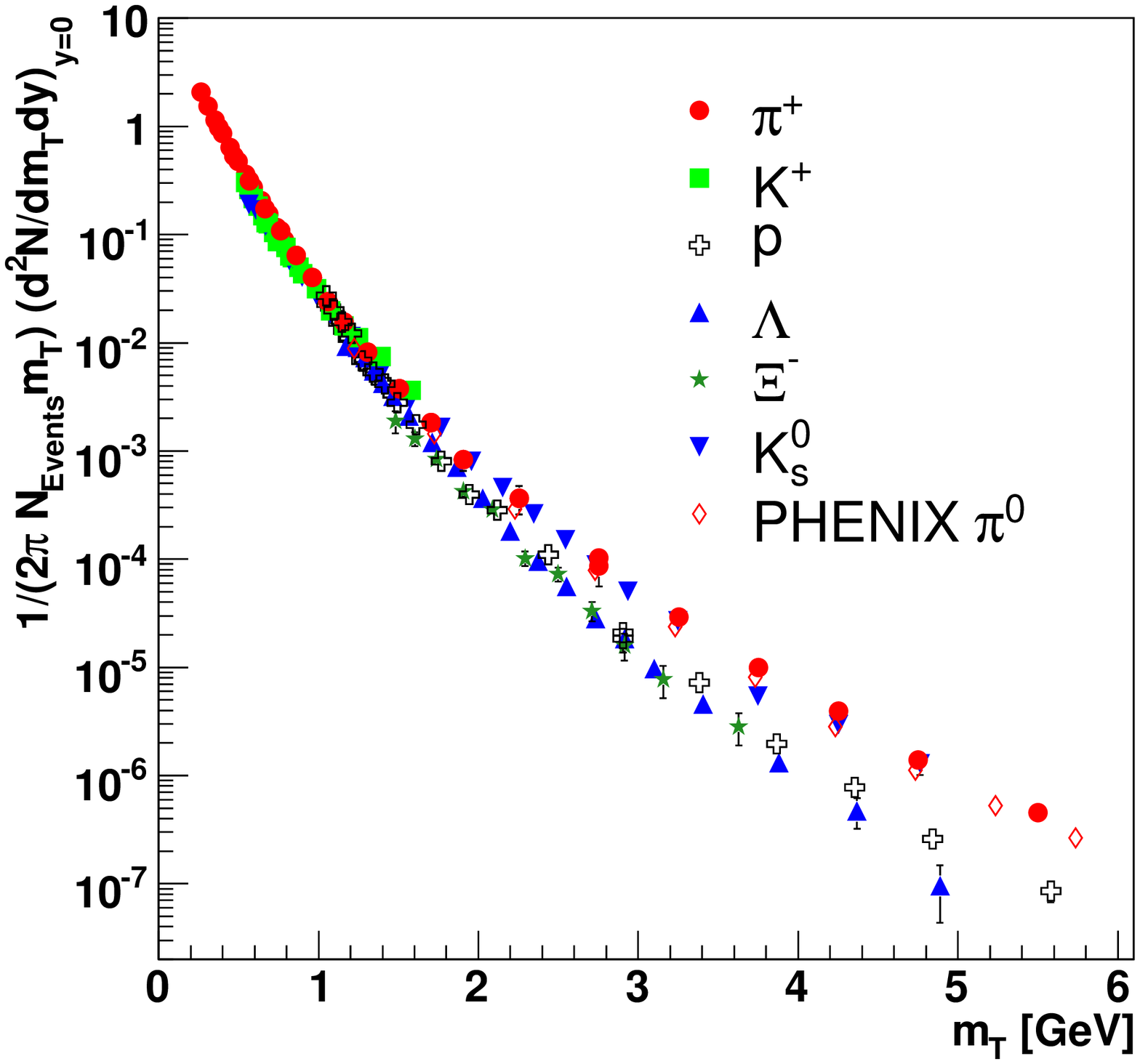, height=4cm, width=4cm}} \quad
\subfigure[PYTHIA
Gluon-jets]{\epsfig{figure=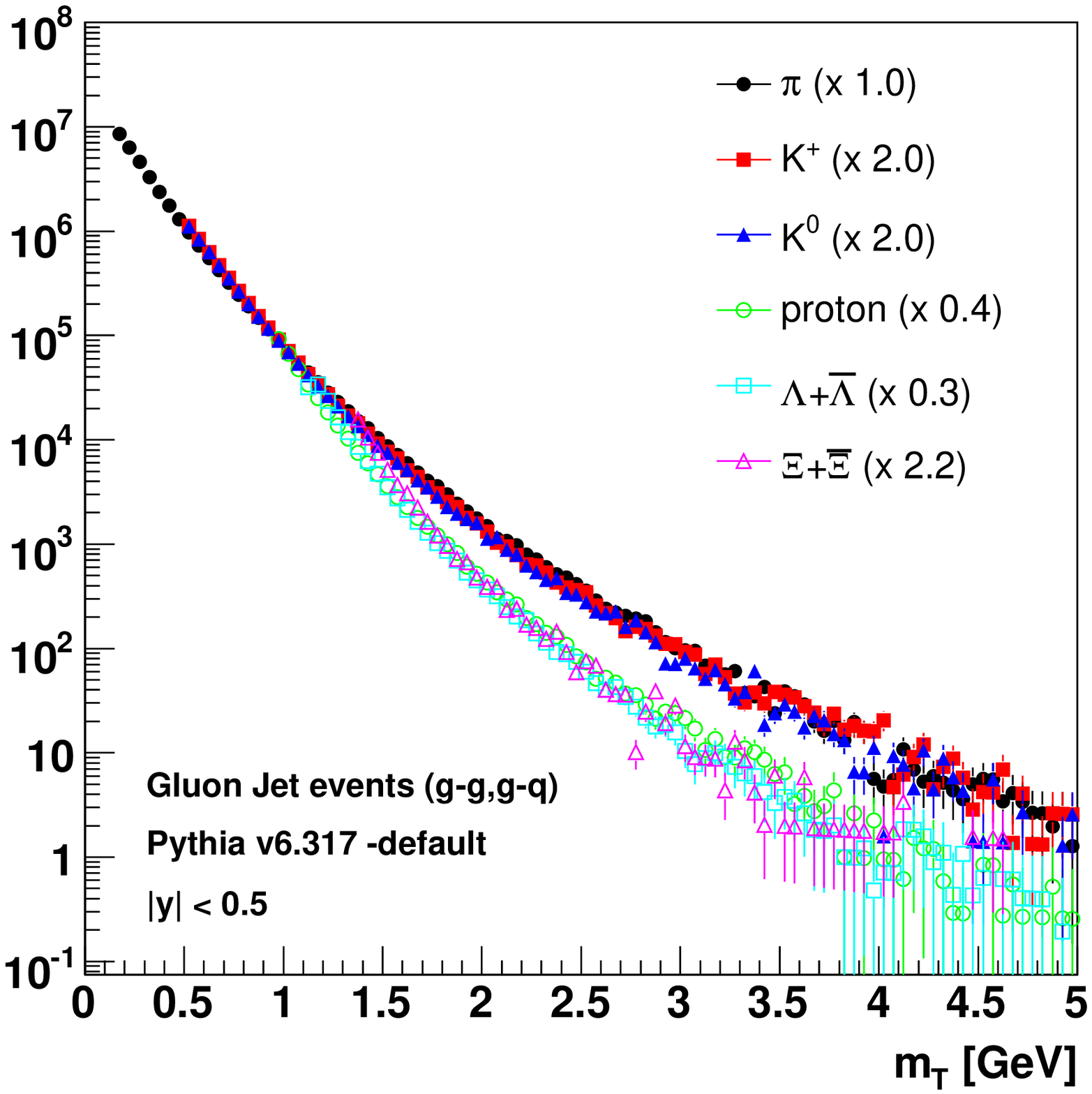, height=4cm,
width=4cm}} \subfigure[PYTHIA
Quark-jets]{\epsfig{figure=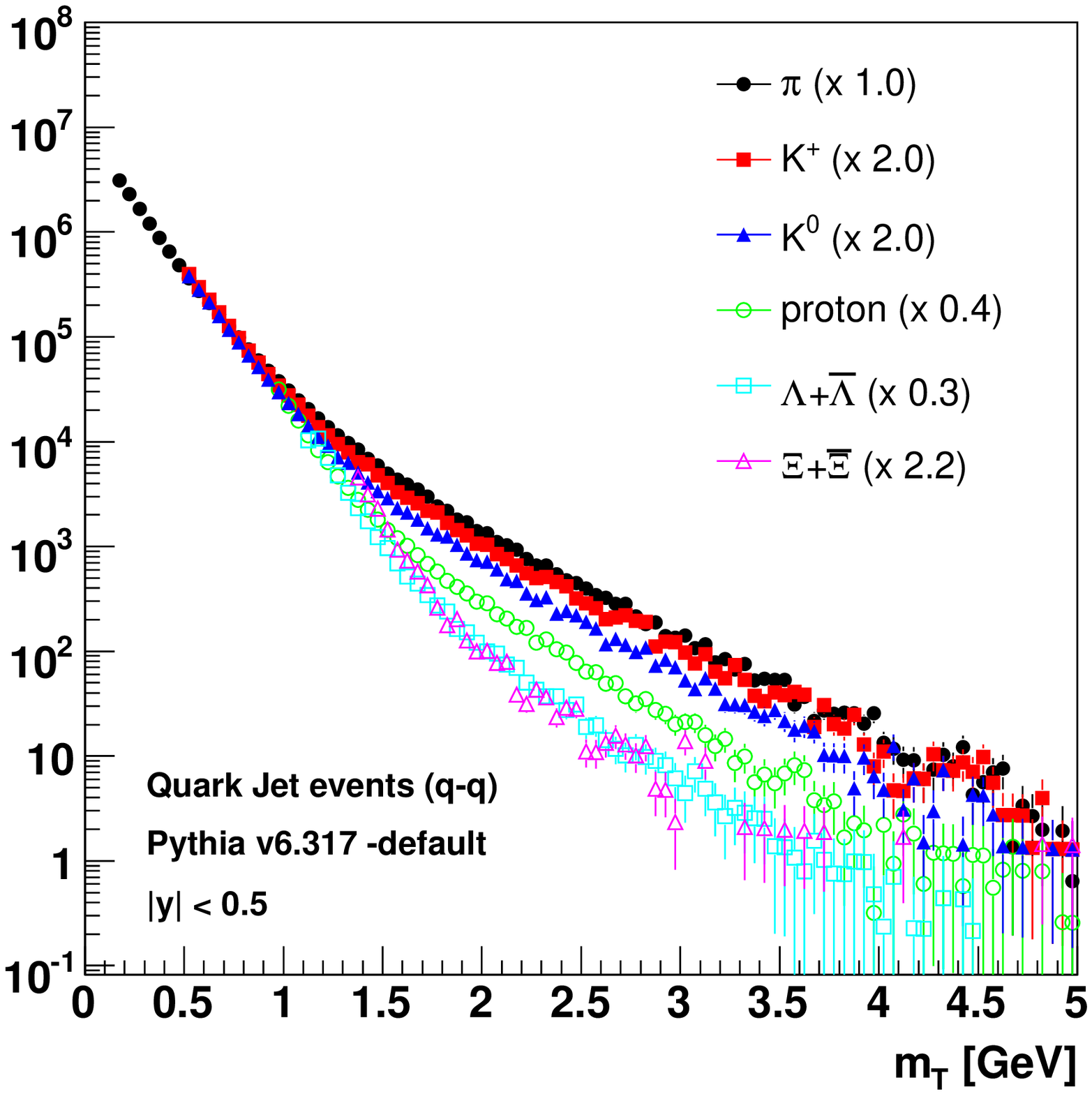, height=4cm,
width=4cm}}}
\caption{(Left)Arbitrarily scaled \mt spectra for
baryons and mesons from $p+p$ collisions at $\sqrt{s} = 200 $GeV.
(center) Scaled \mt~spectra for Gluon-jet events from PYTHIA.
(right) Scaled \mt~spectra for Quark-jet events from PYTHIA.}
\label{fig:mtscaling}
\end{figure}

It is interesting to observe that gluon jets will fragment very
differently into baryons and mesons than quark jets. For gluon jets,
there is a clear shape difference between baryons and mesons at \mt
$\sim$ 1.5GeV, consistent with the di-quark suppression in the
string fragmentation picture. For quark jets, the shape difference
is modified by an additional dependency on mass of the produced
particle. When comparing the result from PYTHIA with our data, this
picture indicates the dominance of gluon jets in \pp at RHIC
energies.

\section{Baryon production in pQCD}

In string models (i.e.PYTHIA) baryon production is understood via
the production of di-quarks pairs from string-breaking and their
recombination with other quarks. This process is suppressed with
respect to $\bar{q}-q$ pairs from string-breaking resulting in
systematically lower baryon yields than mesons.

\begin{figure}[ht]
\centering \mbox{\subfigure[STAR data vs
PYTHIA]{\epsfig{figure=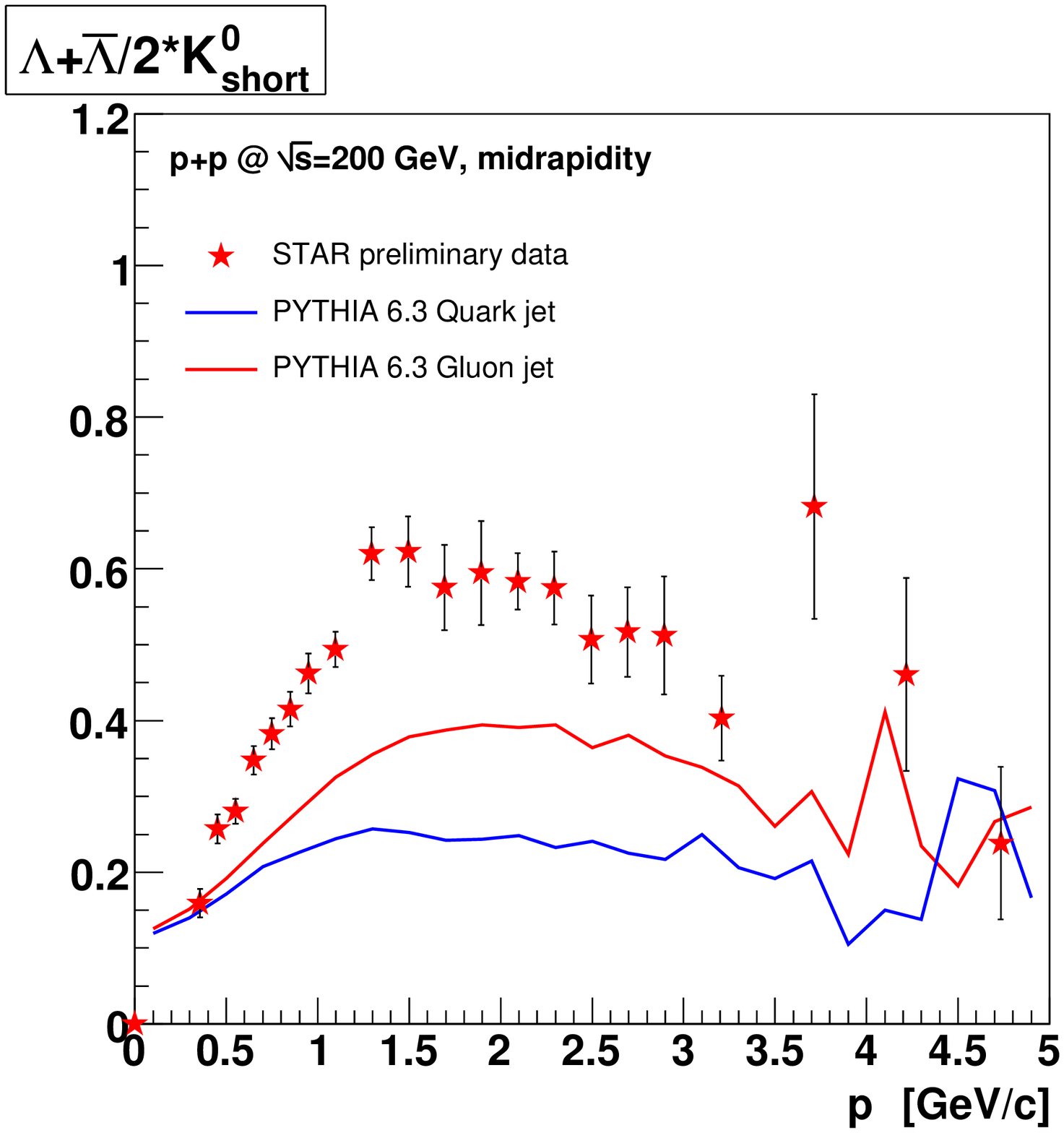, height=5cm,
width=5cm} } \subfigure[STAR vs
UA1]{\epsfig{figure=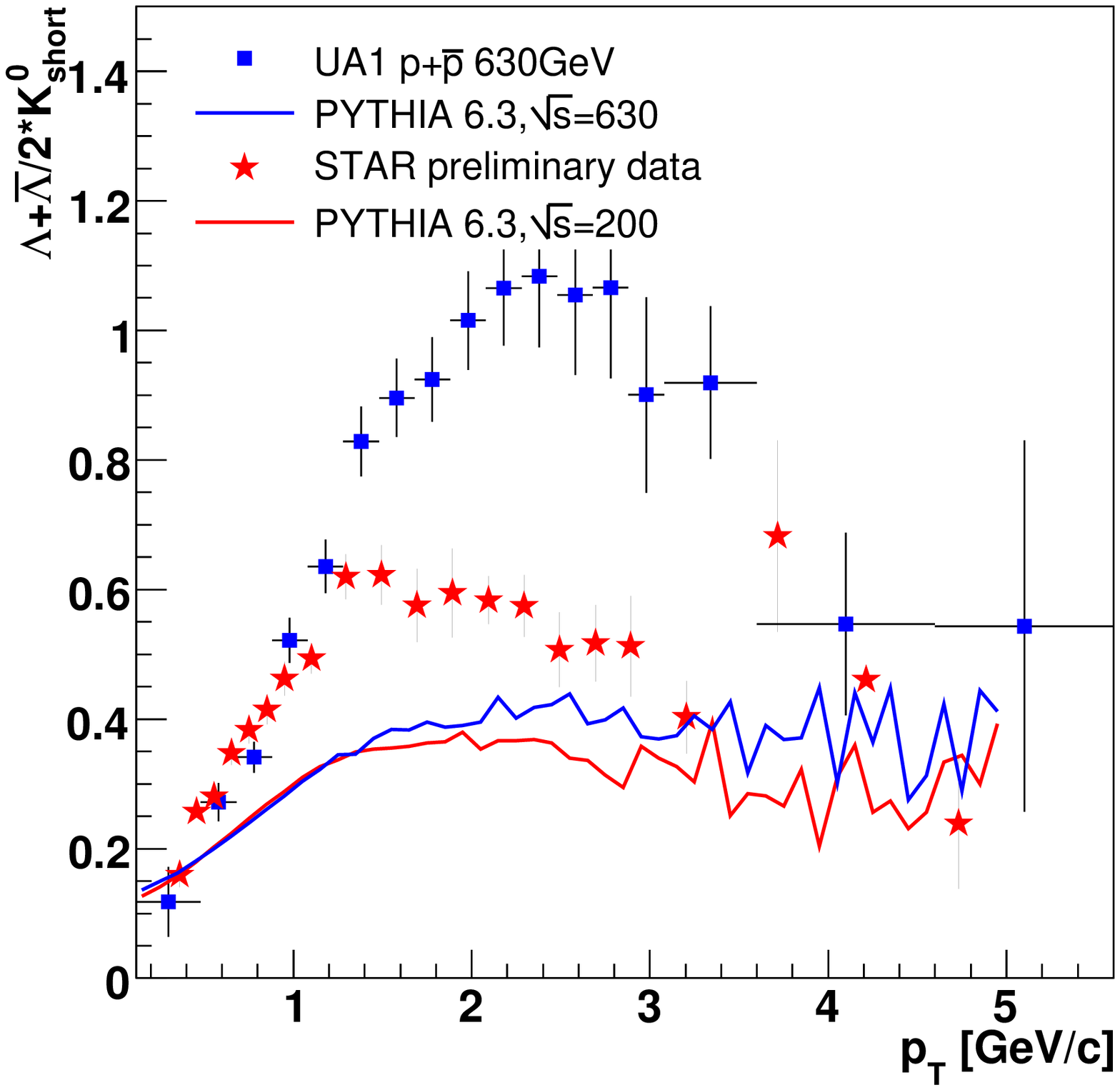, height=5cm, width=5cm}
}} \caption{Ratio of \lam over \kz vs \pt at midrapity from $p+p$ at
$\sqrt{s} = 200 $GeV and $\sqrt{s} = 630$~GeV compared to
predictions from PYTHIA. UA1 data from \cite{UA1:PLB366}. }
\label{fig:LambdaK0s}
\end{figure}

Recent heavy ion data from STAR show a large enhancement of the
baryon to meson ratios at intermediate \pt, which is associated with
parton coalescence and recombination models \cite{Lamont:06}. Figure
\ref{fig:LambdaK0s} (a) shows the predictions from PYTHIA for the
\lam~/\kz ratio vs \pt. We have separated the results for gluon and
quark jet events to show how these contribute to the ratio
differently. The overall PYTHIA result lies in between, closer to
the red line. We observe that the prediction underestimates our data
by at least a factor of 2 in the \pt range from 0.8-3 GeV/c.

In figure \ref{fig:LambdaK0s} (b) we show that the disagreement is
not specific to our energy scale but also exists at Tevatron
energies. At \sqsSps~GeV, the difference between PYTHIA and data is
about a factor of 3 and the enhancement of \lam~/\kz is twice as
large as in STAR. This may be an indication that the effects
observed in this ratio in heavy-ion data are present in some form in
\pp data and therefore this should be noted before attributing
heavy-ion phenomena to explain this effect.

\section{Summary}
We have shown that the theoretical description of identified strange
particles in \pp and \ppbar collisions is still not fully
understood. This is especially important since these models are now
extensively used to predict observables for the LHC-era, and
therefore one should be aware of their limitations. Phenomenological
LO models can be tuned to describe the data but are unable to
describe the baryon enhancement at intermediate \pt. NLO calculation
have greatly improved with light flavor tagged fragmentation
functions. However the high-z range of the gluon FF previously
extracted from \ee data seems inconsistent with \pp and \ppbar data,
indicating that RHIC data could be valuable in constraining the
gluon FF. Arbitrarily scaled \mt~spectra for strange particles
exhibit \mt~scaling and confirm the dominance of gluon jets in \pp
and therefore the importance of understanding gluon fragmentation.
Finally, the baryon to meson ratio at intermediate \pt is about a
factor 2 larger than predicted by pQCD. This difference is even
larger at \sqsSps GeV in \ppbar collisions. This is an indication
that the baryon/meson effects previously observed in heavy ion
collisions are present in some form in \pp data, and that the
associated physics phenomena therefore need to be explained without
requiring the presence of a quark-gluon plasma.

\section*{Acknowledgments}
The author would like to acknowledge theoretical calculations and
enlightening discussions with Simon Albino, Peter Skands and Werner
Vogelsang.

\vfill\eject
\end{document}